\documentclass[useAMS,usenatbib]{mn2e}
\usepackage{aas_macros}
\usepackage[a4paper,centering, totalwidth=520pt, totalheight=700pt]{geometry}
\usepackage[fleqn]{amsmath} 
\usepackage{graphicx}
\usepackage{amssymb}
\usepackage{amsmath}
\usepackage{microtype}
\usepackage{color}

\def\k#1 {k_{{\rm #1}}}

\def\mH2p{H_2^+}

\def\ltsima{$\; \buildrel < \over \sim \;$}
\def\simlt{\lower.5ex\hbox{\ltsima}}   
\def\gtsima{$\; \buildrel > \over \sim \;$}
\def\gtsim{\lower.5ex\hbox{\gtsima}}

\title[Mass modeling of galaxy clusters]{Mass modeling of galaxy clusters: quantifying hydrostatic bias and contribution from non-thermal pressure}
\author[D. Martizzi et al.]{\parbox[t]{\textwidth}{Davide Martizzi$^{1}$\thanks{E-mail: dav.martizzi@berkeley.edu} \& Harrison F. Agrusa$^{1}$\thanks{E-mail: harrisonagrusa@berkeley.edu}} \\
$^{1}$Department of Astronomy and Theoretical Astrophysics Center, University of California, Berkeley, CA 94720-3411, USA.\\ }

\begin{document}

\maketitle

\label{firstpage}

\begin{abstract}
Galaxy cluster mass determinations achieved using X-ray and Sunyaev-Zel'dovich data combined with the assumption of hydrostatic equilibrium are generally biased. The bias exists for two main reasons: non-thermal pressure forces are expected to contribute to the overall pressure balance and deviations from spherical symmetry and hydrostatic equilibrium can be present. In this paper, we use a sample of zoom-in hydrodynamical simulations of galaxy clusters to measure the magnitude of  hydrostatic bias and the non-thermal contribution to the total pressure. We propose a new empirical model for non-thermal pressure based on our simulations that can be applied to observations. We show that our model can be successfully applied to remove most of the bias related to neglection of non-thermal pressure, which is usually not included in hydrostatic cluster mass profile reconstructions. The use of this model may significantly improve the calibration of cluster scaling relations that are a key tool for cluster cosmology. 
\end{abstract}



\section{Introduction}

Galaxy clusters are composite gravitationally bound systems hosting hundreds of galaxies and large quantities of hot X-ray emitting gas (intracluster medium, ICM). Very early work by \cite{1933AcHPh...6..110Z} on the dynamics of galaxies within clusters revealed that these systems host large quantities of dark matter. 

Being the largest virialized structures in the Universe and typically sitting at the knots of the cosmic web, galaxy clusters can be used to trace large scale structure \citep[e.g.][]{1988ARA&A..26..631B}, to measure the abundance of baryonic matter \citep[e.g.][]{2004ApJ...617..879L,2006ApJ...640..691V,2007ApJ...666..147G,2009ApJ...703..982G,2013ApJ...778...14G,2013MNRAS.435.3469H,2013A&A...555A..66L,2016A&A...592A..12E} and  to place constraints on cosmological parameters \citep[e.g.][]{2014A&A...571A..20P, 2014MNRAS.440.2077M,2014arXiv1411.8004H,2014A&A...570A..31B,2016arXiv160306522D}. 

Measurements of the galaxy cluster mass function in the era of large volume surveys (e.g. Euclid, Dark Energy Survey, eROSITA) are expected to  yield cosmological constraints with precision and accuracy at the $\sim 1\%$ level \citep[e.g.][]{2013JCAP...04..022B, 2014MNRAS.441.1769C, Martizzi2014a, 2014MNRAS.439.2485C, 2014A&A...571A..20P, 2016MNRAS.456.2361B}. Accurate measurements of the galaxy cluster mass function require reliable calibrations of scaling relations between the halo mass and observable quantities. Widely used scaling relations are derived from properties measured from clusters detected in the X-ray band and with the thermal Sunyaev-Zel'dovich (SZ) effect \citep[e.g.][]{2010A&A...517A..92A, 2012MNRAS.427.1298H, 2013ApJ...767..116M, 2013ApJ...772...25S, 2013SSRv..177..247G, 2016arXiv160604983D}. Calibration of these scaling relations requires accurate determination of cluster masses.  

A variety of estimators exist for measuring cluster masses, including the use of  gravitational lensing \citep{2013SSRv..177...75H}, X-ray/SZ data combined with the assumption of hydrostatic equilibrium \citep{2013SSRv..177..119E} and the caustic method \citep{2015arXiv151107872M}. Reconstruction of cluster mass profiles under the assumption of hydrostatic equilibrium and spherical symmetry is one of the most common approaches because of its simplicity. However, differences between mass determinations using X-ray data and gravitational lensing are well documented in the literature: it has been argued that deviations from spherical symmetry \citep[e.g.][]{2010ApJ...713..491M, 2014ApJ...794..136D} are a major source of hydrostatic bias. The neglect of pressure support from non-thermal processes such as turbulence in the ICM, bulk motions, magnetic fields, cosmic ray pressure and electron-ion non-equilibrium in the ICM \citep{2007MNRAS.378..385P, 2009ApJ...705.1129L, 2010ApJ...711.1033Z, 2012ApJ...758...74B, 2013ApJ...771..102F, 2013MNRAS.432..404M, 2015ApJ...808..176A} is also thought to contribute to this bias. 

Independent of the method chosen to reconstruct the mass distribution of a cluster, it is important to assess possible biases in the determination of halo masses. If mass determinations for a given cluster sample have been performed with several techniques (e.g. weak lensing and hydrostatic masses from X-ray measurements), observational data can be used to constrain the bias \citep{2015MNRAS.450.3633S}. 
On the theoretical side, it is possible to use numerical cosmological simulations of galaxy clusters to directly compare the actual mass profile of a simulated cluster with the one obtained via reconstruction using the methods cited above \cite[e.g.][]{2013ApJ...777..151L}. Driven by the increasing relevance of mass calibration needed for cluster surveys, several groups pursued the goal of constraining non-thermal pressure support to improve hydrostatic mass determinations. In the literature, particular emphasis has been given to the contribution of turbulent and bulk motions induced by  active galactic nuclei (AGN) feedback, sloshing, and mergers \citep{2012A&A...544A.103V}. In a series of papers,  \cite{2014MNRAS.442..521S}, \cite{2015MNRAS.448.1020S} and \cite{2016MNRAS.455.2936S} developed an analytical model for non-thermal pressure from turbulent and bulk motions in galaxy clusters and used it remove hydrostatic bias; however, this model is based on knowledge of the gas velocity dispersion which is complicated to measure experimentally. The recent paper by \cite{2016arXiv160602293B} used cosmological SPH simulations to characterize hydrostatic bias in galaxy clusters but did not directly implement a method to reconstruct the mass profile. The goal of our paper is to extend this line of research by using a sample of cosmological adaptive mesh refinement (AMR) simulations of galaxy clusters to constrain hydrostatic bias, to quantify the role of non-thermal pressure and to develop a new method for mass reconstruction. 
Furthermore, we provide novel insight on the non-thermal pressure in cluster cores in AMR simulations that include AGN feedback. 

The paper has the following structure. Section 2 describes the simulation data-set. Section 3 reviews the formalism of hydrostatic mass modeling, describes the analysis procedure and proposes an empirical model for the non-thermal pressure. Section 4 describes our results. Finally, Section 5 summarizes our findings.

\section{Simulation data-set}
\label{sec:simulation_dataset}

We consider a sub-set of 10 cosmological hydrodynamical zoom-in simulations of
galaxy clusters from the sample of \cite{Martizzi2014a}. These simulations were performed with the {\scshape ramses} code \citep{2002A&A...385..337T} and were shown to produce brightest cluster galaxies (BCGs) with realistic mass, sizes and kinematic properties at redshift $z=0$ \citep{Martizzi2014b}, and the host halos have baryon and stellar fractions in broad agreement with observations \citep{Martizzi2014a} and have realistic ICM density and thermal pressure profiles (as we show in Subsection~\ref{sec:measured_quantities}). The zoom-in simulations of the 10 halos considered in this paper have {\it total} masses greater than $10^{14}$~M$_\odot$. Furthermore, neighbouring halos do not have masses larger than half of the total central halo mass within a spherical region of five times the virial radius of the central halo. Half of the clusters in the sample used for this paper are relaxed, based on the criterion described in Subsection~\ref{sec:relaxation}. 

\begin{table}
\begin{center}
{\bfseries Cosmological parameters}
\begin{tabular}{|c|c|c|c|c|c|}
\hline
\hline
 ${\rm H_0}$ [km s$^{-1}$Mpc$^{-1}$] & ${\rm \sigma_{\rm 8}}$ & ${\rm n_{\rm s}}$ & $\Omega_\Lambda$ & $\Omega_{\rm m}$ & $\Omega_{\rm b}$ \\
\hline
\hline
 70.4 & 0.809 & 0.963 & 0.728 & 0.272 & 0.045 \\
\hline
\hline
\end{tabular}
\caption{Cosmological parameters adopted in our simulations. }\label{tab:cosm_par}
\end{center}
\end{table}

Cosmic structure formation is evolved in the context of the standard $\Lambda$CDM cosmological scenario. The cosmological parameters chosen for our simulations are summarised in Table~\ref{tab:cosm_par}. Cosmological initial conditions have been considered which were computed using the \cite{1998ApJ...496..605E} transfer function and the {\scshape grafic++} code developed by Doug Potter (http://sourceforge.net/projects/grafic/) and 
based on the original {\scshape grafic} code \citep{2001ApJS..137....1B}. 

The {\scshape ramses} code offers AMR capabilities which is fully exploited to achieve high resolution in our zoom-in simulations. The computational domain is a box size $144$~Mpc/h. We chose an initial level of refinement $\ell=9$ (base mesh size $512^3$), but we allowed for refinement down to a maximum level $\ell_{\rm max}=17$. Grid refinement is implemented using a quasi-Lagrangian approach: when the dark matter or baryonic mass in a cell reaches 8 times the initial mass resolution, it is split into 8 children cells. With these choices, the dark matter particle mass is ${\rm m_{\rm cdm}=1.62\times 10^{8}}$~M$_\odot$/h, while the mass of the baryon resolution element is ${\rm m_{\rm gas}=3.22\times 10^{7}}$~M$_\odot$. The minimum cell size in the zoom-in region is ${\rm \Delta x_{\rm min} = L/2^{\ell_{\rm max}}\simeq 1.07}$ kpc/h. Table~\ref{tab:mass_par} summarizes the particle mass and spatial resolution achieved in our simulations.

\begin{table}
\begin{center}
{\bfseries Mass and spatial resolution}
\begin{tabular}{|c|c|c|}
\hline
\hline
${\rm m_{\rm cdm}}$&  ${\rm m_{\rm gas}}$ & ${\rm \Delta x_{\rm min}}$ \\
$[10^{8}$ M$_\odot$/h] & $[10^{7}$ M$_\odot$/h] & [kpc/h] \\
\hline
\hline
 $1.62$ & $3.22$ & $1.07$ \\
\hline
\hline
\end{tabular}
\end{center}
\caption{Mass resolution for dark matter particles, gas cells and star particles, and spatial resolution (in physical units) for our simulations. }\label{tab:mass_par}
\end{table}

Our zoom-in simulations include baryons (hydrodynamics), star formation, stellar feedback and AGN feedback. The inviscid equations of hydrodynamics are solved with a second-order unsplit Godunov scheme \citep{2002MNRAS.329L..53B, Teyssier:2006p413, Fromang:2006p400} based on the HLLC Riemann solver and the
MinMod slope limiter \citep{Toro:1994p1151}. We assume a perfect gas equation of state (EOS) with polytropic index $\gamma=5/3$. All the zoom-in runs include sub-grid models for gas cooling which account for H, He and 
metals and that use the \cite{1993ApJS...88..253S} cooling function. We directly follow star formation and supernovae feedback (``delayed cooling" scheme, \citealt{2006MNRAS.373.1074S}) and metal enrichment. 
The AGN feedback scheme is a modified version of the \cite{2009MNRAS.398...53B} model. Supermassive black holes (SMBHs) are modeled as sink particles and AGN feedback is provided in form of thermal energy injected in a sphere surrounding each SMBH. More details about the AGN feedback scheme can be found in \cite{Martizzi2012a}.

\subsection{Cluster classification by relaxation state}\label{sec:relaxation}

The classification of clusters depending on their relaxation state is critical for our analysis. Reliable mass reconstructions from observational data are usually performed by excluding clusters that are not dynamically relaxed. 
In this paper, we also analyze the dynamical state of the simulated clusters using a simple estimator that has been shown to be reliable by \cite{2016MNRAS.tmp.1545C}. 

At each cluster-centric distance ${\rm r}$, we define the virial parameter $\eta(<r)$ as
\begin{equation}
 {\rm \eta(<r) = \frac{\sigma_{\rm dm}(<r)}{\sigma_{\rm vir}(<r)}},
\end{equation}
where ${\rm \sigma_{\rm dm}(<r)}$ is the velocity dispersion of dark matter particles within radius ${\rm r}$, and ${\rm \sigma_{\rm vir}(r)}$ is the velocity dispersion expected if the system were virialized within the same radius, i.e.
\begin{equation}
 {\rm \sigma_{vir}(<r)=\left[\frac{GM_{\rm dm}(<r)}{r}\right]^{1/2}}
\end{equation}
where ${\rm M_{\rm dm}(<r)}$ is the enclosed dark matter mass. Any deviation from $\eta(<r)=1$ implies deviations from virial equilibrium within radius ${\rm r}$. 

\begin{figure}
\includegraphics[width=.49\textwidth]{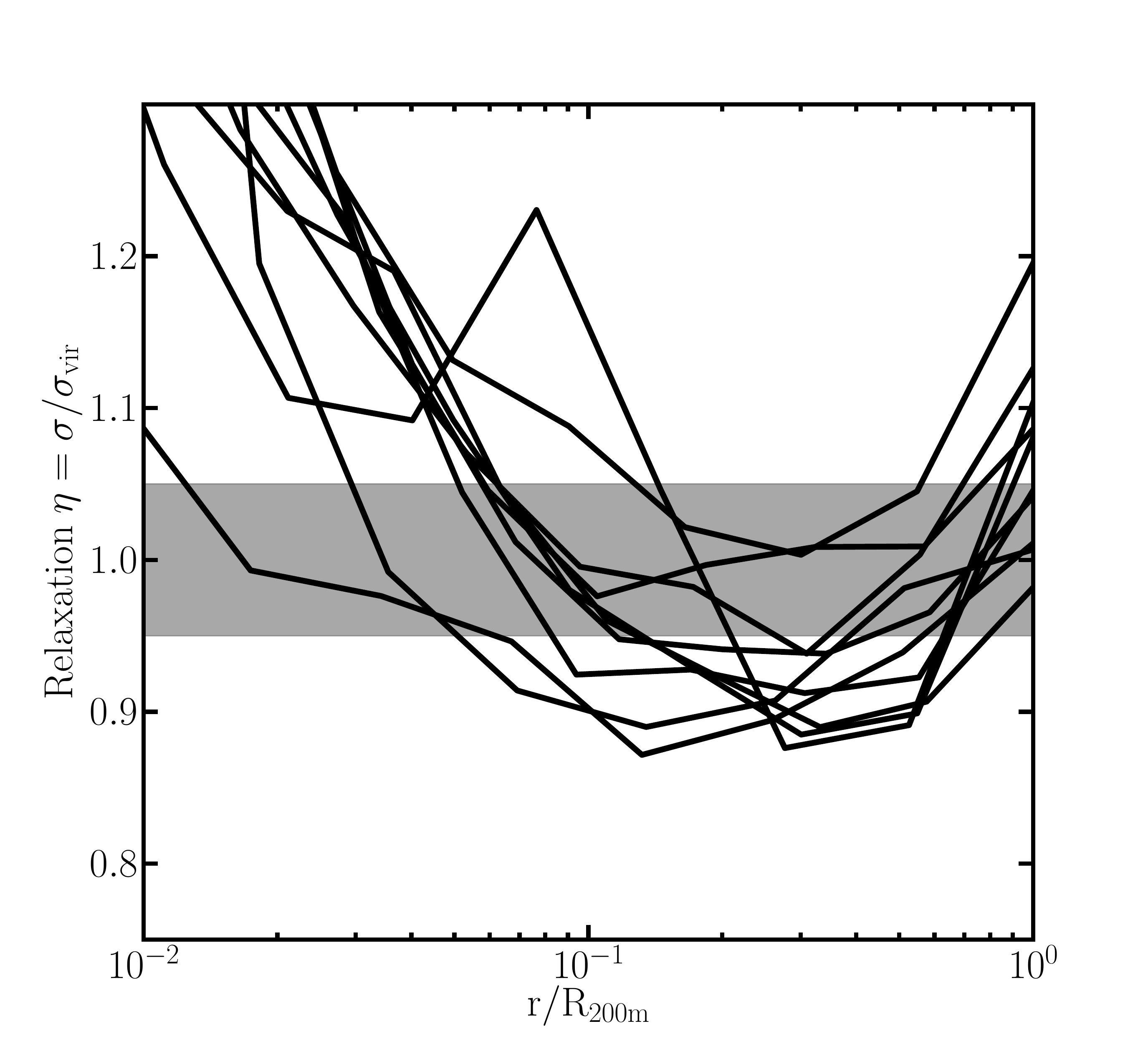}
\caption{\label{relaxation} Relaxation parameter $\eta(<r)=\sigma(<r)/\sigma_{\rm vir} (<r)$ within a given radius $r$, where $\sigma$. Each line represents the relaxation profile of one of the simulated clusters. $\eta=1$ is achieved only at full virialization. 
The shaded area represents $\pm5\%$ deviations from pure virialization. } 
\end{figure}

There are several caveats that one should consider when using such measure of relaxation. First of all, ${\rm \eta(<r)}$ is expected to vary with radius. A cluster may appear to be relaxed within a certain radius ${\rm r_1}$, but not within another radius ${\rm r_2}$. 
Figure~\ref{relaxation} shows that ${\rm \eta(<r)}$ varies significantly as a function of radius for all our clusters. In particular, the strongest deviations from virialization are observed in cluster cores (${\rm r<0.1R_{200m}}$) and at intermediate radii, as a consequence of the 
presence of non-virialized sub-structure in the cluster. The second caveat is that we have only used dark matter to define ${\rm \eta}$ (for simplicity), but the dynamics in cluster cores might be significantly influenced by the baryonic mass/velocity distribution.

To overcome the limitations of the definition of ${\rm \eta}$, we place a series of restrictive requirements when classifying clusters depending on their dynamical state. We define a cluster as relaxed only if the two following requirements are simultaneously satisfied:
\begin{itemize}
 \item ${\rm 0.95\leq \eta(<R_{200m}) \leq 1.05}$.
 \item ${\rm 0.90\leq \eta(<0.5R_{200m}) \leq 1.10}$.
\end{itemize}
In other words, a cluster is significantly relaxed only if its virial parameters at ${\rm 0.5R_{200m}}$ and ${\rm R_{\rm 200m}}$ do not deviate significantly from 1. We find that 5 of our halos are relaxed (labeled with IDs from 1 to 5), whereas the other 5 clusters are unrelaxed (labeled with IDs from 6 to 10).

\section{Formalism}\label{sec:formalism}

The mass distribution of a galaxy cluster can be inferred from observational measurements via simplifying assumptions about the state of the cluster. If the cluster ICM is modeled as a fluid in the potential generated by the total mass distribution, then the gas motions are described by Euler's equation: 
\begin{equation}\label{eq:euler}
\frac{\rm d {\bf v}}{\rm dt} = {\rm -\nabla \Phi} - \frac{\rm \nabla P}{\rho}.
\end{equation}
where ${\rm {\bf v}} $ is the gas velocity, $P$ is the local pressure, $\rho$ is the gas density and $\Phi$ is the gravitational potential. $\Phi$ is determined by solving Poisson's equation:
\begin{equation}
{\rm \nabla^2\Phi= 4\pi G \rho_{total}}
\end{equation}
where ${\rho_{\rm total}}$ is the total mass density in the cluster (dark matter, stars, ICM, etc.). 

Euler's equation can be simplified into a form that leads to a widely used galaxy cluster mass estimator under the assumption of {\it spherical symmetry} and {\it hydrostatic equilibrium} ($\rm d{\bf v}/dt =0$). In this case Equation~\ref{eq:euler} becomes:
\begin{equation}\label{eq:hse}
\frac{\rm d P_{\rm HE}}{\rm dr} = {\rm g_r \rho(r)} = -{\rm \frac{GM_{HE}(<r)\rho(r)}{r^2}},
\end{equation}
where ${\rm r}$ is the cluster-centric distance, ${\rm g_r=-d\Phi/dr}$ is the radial component of the gravitational acceleration, $\rm M_{HE}(<r)$ is the total mass enclosed within radius $r$; we have also introduced $\rm P_{\rm HE}(r)$, the pressure profile required for the cluster to maintain hydrostatic equilibrium. 

If the pressure gradient ${\rm dP_{HE}/dr}$ and density profile $\rm \rho(r)$ are known, then Equation~\ref{eq:hse} can be solved for $\rm M_{HE}(<r)$, yielding an estimate of the cluster mass profile as a function of radius. However, $\rm M_{HE}(<r)$ is an accurate estimate of the total mass profile only if hydrostatic equilibrium and spherical symmetry are good approximations. In general, the real mass profile ${\rm M_{real}(<r)\neq M_{HE}(<r)}$, i.e. the hydrostatic mass is a biased estimator. This effect, hydrostatic bias, can be quantified by measuring ${\rm M_{HE}(<r)/M_{real}(<r)}$ in simulated clusters like the ones considered in this paper.

Despite the existence of possible hydrostatic bias, there is an additional bias that derives from the modeling of the source of pressure support in Equation~\ref{eq:hse}. In the most general sense $\rm P_{\rm HE}$ represents the effect of several physical processes that provide pressure support to the cluster against its own gravity. Contribution to $\rm P_{HE}$ may come from thermal pressure of the ICM, $\rm P_{therm}$, turbulent and bulk motions acting as an effective pressure \cite[e.g.][]{2014MNRAS.442..521S} and other non-thermal processes such as magnetic fields, cosmic ray pressure, and electron-ion non-equilibrium in the ICM. To keep our formalism as general as possible, we label these contribution as {\it non-thermal pressure} $\rm P_{nt}$. Under this assumption we have 
\begin{equation}
{\rm P_{HE} = P_{therm} + P_{nt}}.
\end{equation}
If $\rm P_{nt}$ is significantly large and is intentionally neglected, then Equation~\ref{eq:hse} will yield a biased mass estimate even if perfect hydrostatic equilibrium holds. 

If only thermal pressure is considered for the mass reconstruction, Equation~\ref{eq:hse} yields a commonly used mass estimator:
\begin{equation}\label{eq:mtherm}
{\rm M_{\rm therm} (<r) = -\frac{r^2}{G\rho(r)}\frac{\rm d P_{\rm therm}}{\rm dr} }.
\end{equation}
The thermal pressure profile can be measured or inferred from X-ray and SZ observations, making this estimator easy to apply to observational data-sets. However, this mass estimator is in principle affected by hydrostatic bias and by the neglection of non-thermal pressure; both these biases need to be properly quantified.

\subsection{Measured quantities}\label{sec:measured_quantities}

Cosmological hydrodynamical simulations can help quantify hydrostatic bias and the contribution from non-thermal pressure. Our simulations do not include magnetic fields and  cosmic ray physics, therefore the only source of non-thermal pressure that we are able to constrain is from turbulent and bulk motions in the ICM. Distinguishing between the contributions from turbulence and bulk motions is beyond the scope of this paper and we limit our analysis to the net effect of both processes. 
Potential caveats and limitations of our simulations are discussed in Section~\ref{sec:conclusion}.

First, we measure the density profile ${\rm \rho (r)}$ and thermal pressure profile ${\rm P_{therm}(r)}$ of each simulated cluster at redshift $z=0$. These profiles are obtained by volume-weighted averaging over spherical shells. The use of volume-weighting smooths out large fluctuations in the profiles that would otherwise be present if mass-weighting were used. Having smoother profiles improves the accuracy of the numerical integration of the differential equations we discuss below. 
We also tested mass-weighted averaging, excluding cells whose density is 2$\sigma$ away from the spherically-averaged density at the same radius (similarly to \citealt{2014ApJ...792...25N}), and found that the measured profiles do not differ more than $\sim 1-2\%$ from the volume-weighted profile. After measuring ${\rm \rho (r)}$ and ${\rm P_{therm}(r)}$, Equation~\ref{eq:mtherm} is used to compute the thermal mass profile ${\rm M_{therm}(<r)}$. 

\begin{figure*}
\includegraphics[width=.99\textwidth]{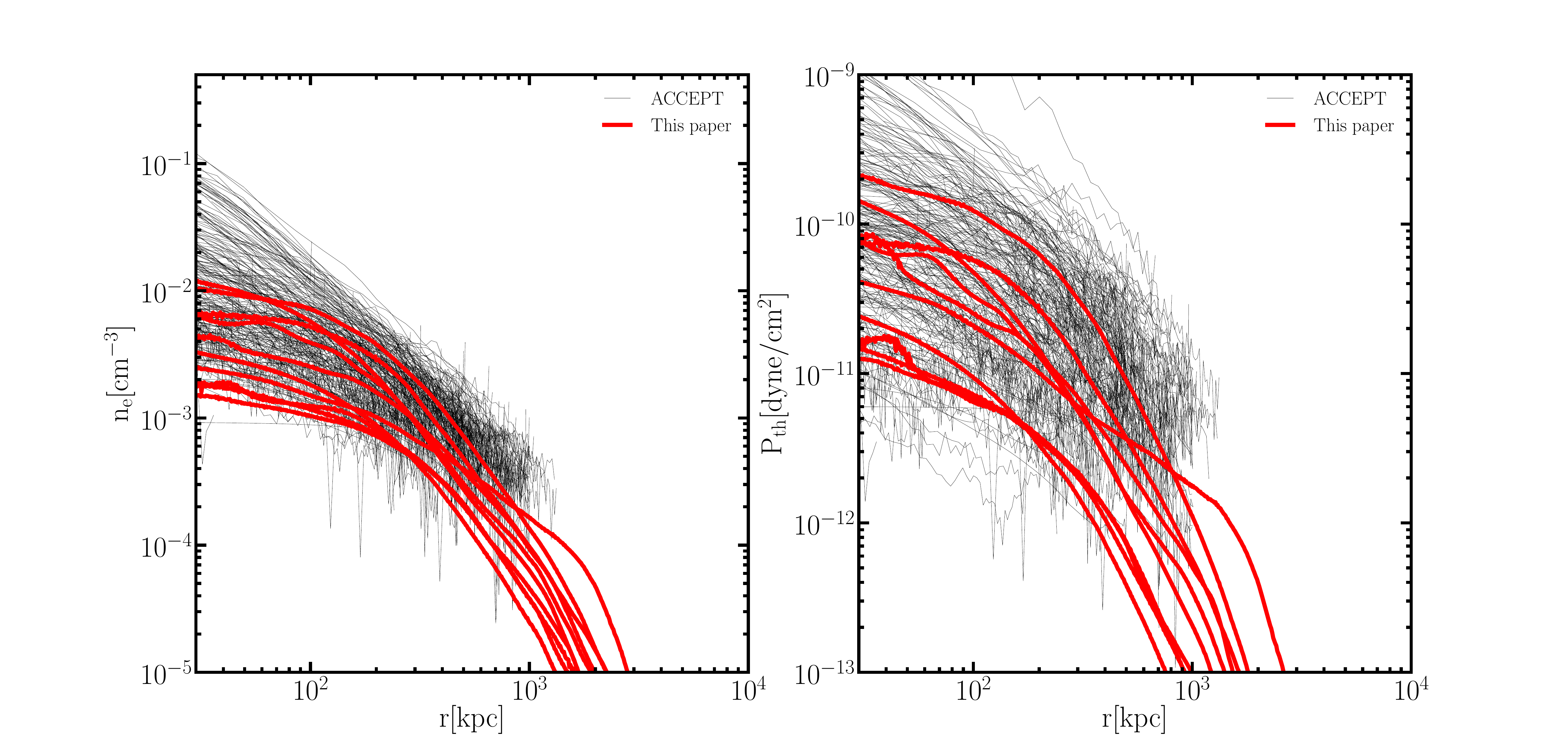}
\caption{\label{rho_P_vs_ACCEPT} Electron number density (left) and thermal pressure (right) profiles from our simulations (red lines) vs. the observed clusters in the ACCEPT sample (Cavagnolo et al. 2009).}
\end{figure*}

Figure~\ref{rho_P_vs_ACCEPT} shows that we achieve satisfactory agreement between the density and thermal pressure profiles in our simulations at redshift $z=0$ and those measured from the ACCEPT cluster sample \citep{2009ApJS..182...12C}. 

According to Equation~\ref{eq:hse}, the total hydrostatic mass of the simulated clusters can be measured as: 
\begin{equation}\label{eq:mass_hse}
{\rm M_{HE}(<r) = -\frac{r^2}{G}g_r},
\end{equation}
where $\rm g_r$ is directly measured from the simulations by projecting the average gravitational acceleration of each AMR cell along the radial direction and subsequently averaging over spherical shells. Due to contribution from non-thermal pressure, ${\rm M_{HE}>M_{therm}}$. 

Subsequently, we reconstruct the total pressure required for hydrostatic equilibrium by numerically integrating Equation~\ref{eq:hse}:
\begin{equation}\label{eq:ptot}
{\rm P_{HE} (r) = P_{HE}(0) - \int_0^r \frac{GM_{HE}(<r')\rho(r')}{r'^2}dr'}
\end{equation}
where we imposed the boundary condition ${\rm P_{HE} (0) = P_{therm}(0)+P_{nt}(0)}$, where the central thermal and non-thermal pressures are measured explicitly (Subsection~\ref{sec:ntp-vdisp}). Under the assumption of hydrostatic equilibrium, the non-thermal pressure profile is simply given by ${\rm P_{nt}(r) = P_{HE}(r) - P_{therm}(r)}$. 

\subsection{Analytical model for non-thermal pressure}

The procedure described in Subsection~\ref{sec:measured_quantities} by which the non-thermal pressure is measured by subtracting the thermal pressure from the total pressure cannot obviously be applied to real systems where in the most ideal scenario only the temperature (pressure) and density profiles of the ICM are known. 

We used our cosmological simulations to calibrate an empirical relation to model the non-thermal pressure as a function of ICM density. We first measure the gas density profile ${\rm \rho(r)}$ and the non-thermal pressure profile ${\rm P_{nt}(r)}$ (see Section~\ref{sec:results}) for all halos. By defining ${\rm x=r/R_{200m}}$, we can express the density 
and non-thermal pressure profile as ${\rm \rho(x)}$ and ${\rm P_{nt}(x)}$, respectively. A point in the ${\rm (\rho,P_{\rm nt})}$ space is associated to each halo and radial position ${\rm x}$. We found that values of ${\rm P_{nt}(x)}$ are correlated to values of ${\rm \rho(x)}$ in the ${\rm (\rho,P_{\rm nt})}$ space. Our analytical model is a fit to the 
empirical correlation we find in our sample with parameters found by performing a standard $\chi^2$ regression. In this model, the non-thermal pressure is given by:
\begin{eqnarray}\label{eq:model_pnt}
 {\rm P}_{\rm nt,model}(r) = 5.388 \times 10^{13}  \times \left(\frac{\rm R_{200m}}{1 \hbox{ Mpc}}\right)^3 \times \nonumber  \\ 
	\times \left[\frac{\rm \rho(r)}{1 \hbox{ g/cm}^3}\right] \hbox{ erg/cm}^3.
\end{eqnarray}

\begin{figure}
\includegraphics[width=.49\textwidth]{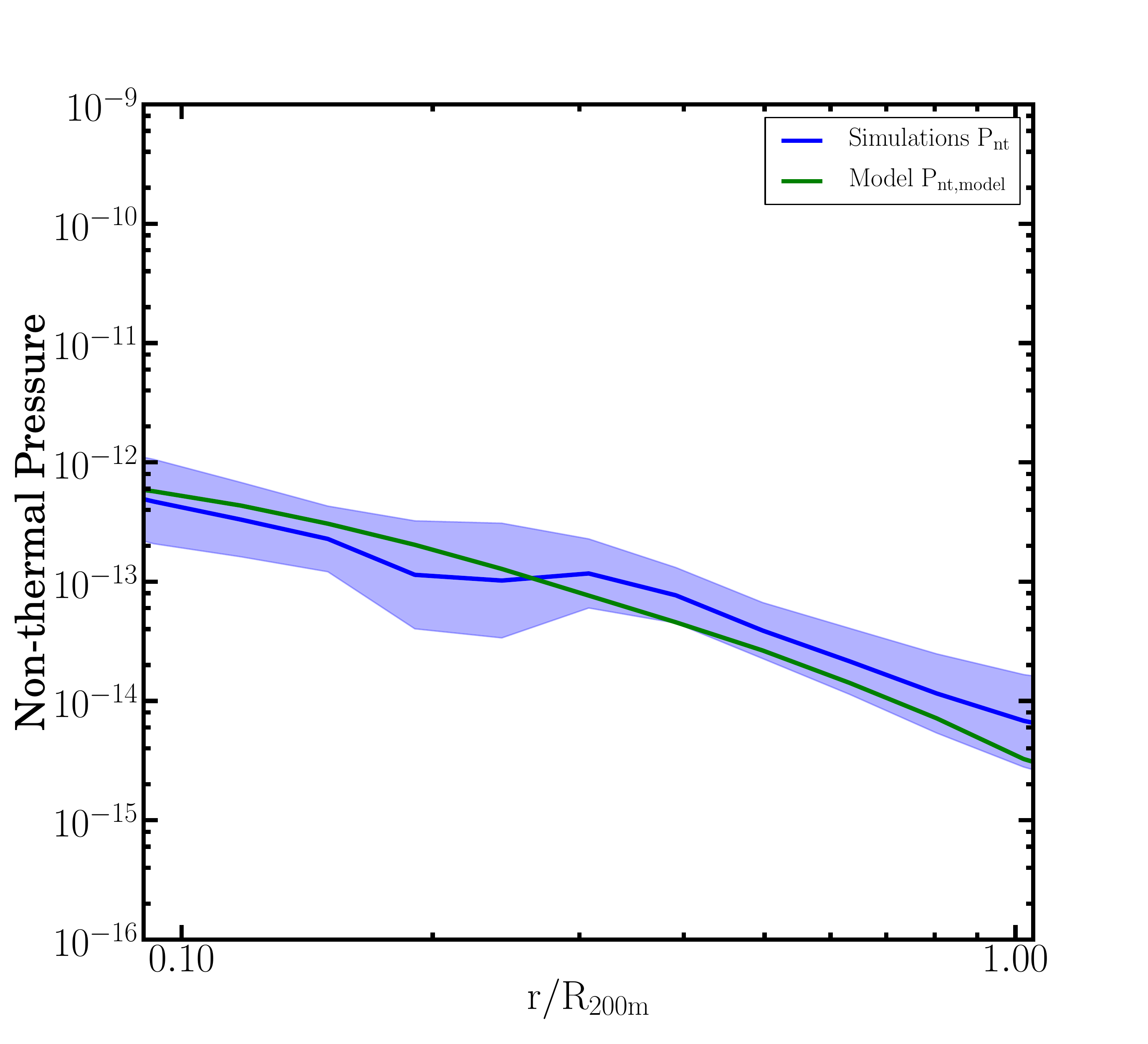}
\caption{\label{fig:ntp} Comparison of the mean non-thermal pressure profile for the relaxed clusters (blue) to the analytical model of Equation~\ref{eq:model_pnt} (green). 
The blue shaded region represents the halo-to-halo 1-$\sigma$ scatter. } 
\end{figure}

Figure~\ref{fig:ntp} shows the mean non-thermal pressure profile measured from the sub-sample of 5 relaxed clusters (blue line, with its 1$\sigma$ halo-to-halo scatter) compared to the analytical model from Equation~\ref{eq:model_pnt} (green line). The analytical model is well within the 1$\sigma$ scatter at all radii ${\rm r<R_{200m}}$. 

Notice how the non-thermal pressure in Equation~\ref{eq:model_pnt} is expressed as a function of ${\rm R_{200m}}$. In recent papers by \cite{2014ApJ...792...25N}, \cite{2015ApJ...806...68L} and \cite{2016arXiv160501723A}, ${\rm R_{200m}}$ was confirmed to be a natural radius to re-scale density, pressure and temperature profiles for halos of different halo masses and at different redshifts. 
In particular, \cite{2014ApJ...792...25N} also provide a fitting formula for the non-thermal pressure in simulations without cooling and star formation that depends on $r/R_{\rm 200m}$ and that provides an excellent fit to simulations up to redshift $z=1.5$. \cite{2016MNRAS.455.2936S} also explore a set of analytical models to accurately describe the total pressure profiles from 
simulated clusters. However, our formula is less complicated and can be used to provide a physical interpretation to our results. Since ${\rm R_{200m}=(M_{200m}/4\pi\bar{\rho}_{200m})^{1/3}}$, where ${\rm \bar{\rho}_{200m} = 200\Omega_m\rho_{crit}}$ is the average matter density within ${\rm R_{\rm 200m}}$, we have:
\begin{equation}
 {\rm P_{\rm nt,model}(r) \propto M_{200m} \frac{\rho(r)}{\bar{\rho}_{200m}}},
\end{equation}
i.e. the non-thermal pressure scales linearly with local fluctuations of the density with respect to the mean density within ${\rm R_{200m}}$. Of course, this is only an average effect, since we are considering spherically averaged quantities. However, it is important to stress that correlations between local density fluctuations and turbulent velocity fluctuations have already 
been discussed in the literature \citep[e.g.][]{2014ApJ...788L..13Z}.

Equation~\ref{eq:model_pnt} can be used for cluster mass reconstruction including non-thermal pressure under the assumption of hydrostatic equilibrium. The resulting mass estimator is:
\begin{equation}\label{eq:mass_theory}
{\rm M_{model} (<r) = -\frac{r^2}{G\rho(r)}\frac{d}{dr}[P_{therm} (r) + P_{nt,model}(r)]}.
\end{equation}

The greatest advantage of the empirical model we propose is that the non-thermal pressure scales linearly with the local ICM density, which can be determined from observations. The dependence on $\rm R_{200m}$ is in practice a factor that allows self-similar scaling of the non-thermal pressure profiles of different clusters. This behaviour is of course only approximate. The dependence ${\rm P_{nt,model}\propto R_{200m}^3}$ is somewhat strong, however ${\rm R_{200m}}$ varies only by a factor $\sim 2$ in the range ${\rm 14 \lesssim \log_{10} (M_{200m}/M_{\odot}) \lesssim 15}$, making the model less sensitive to such variations. 

When applying Equation~\ref{eq:model_pnt} to observations, a sufficiently accurate estimate for ${\rm R_{200m}}$ can be obtained by a simple  iterative algorithm:
\begin{enumerate}
\item Initial condition. From measured $\rm \rho(r)$, $\rm P_{therm}(r)$, determine ${\rm M_{therm}(<r)}$ and use it to make an initial guess for ${\rm R_{200m}}$.
\item Start iterations. From measured $\rm \rho(r)$, $\rm P_{therm}(r)$ and guess for $\rm R_{200m}$, determine $\rm M_{model} (<r)$.
\item Determine new value for $\rm R_{200m}$ and iterate from point number (ii). Repeat until convergence.
\end{enumerate}

In Section~\ref{sec:results} we will quantitatively asses the quality of this approximation comparing ${\rm M_{model}(<r)}$ to the real mass profile of our simulated clusters ${\rm M_{real}(<r)}$. This comparison will highlight the efficiency of this model at removing biases in the determination of cluster mass. 

\begin{figure}
\includegraphics[width=.49\textwidth]{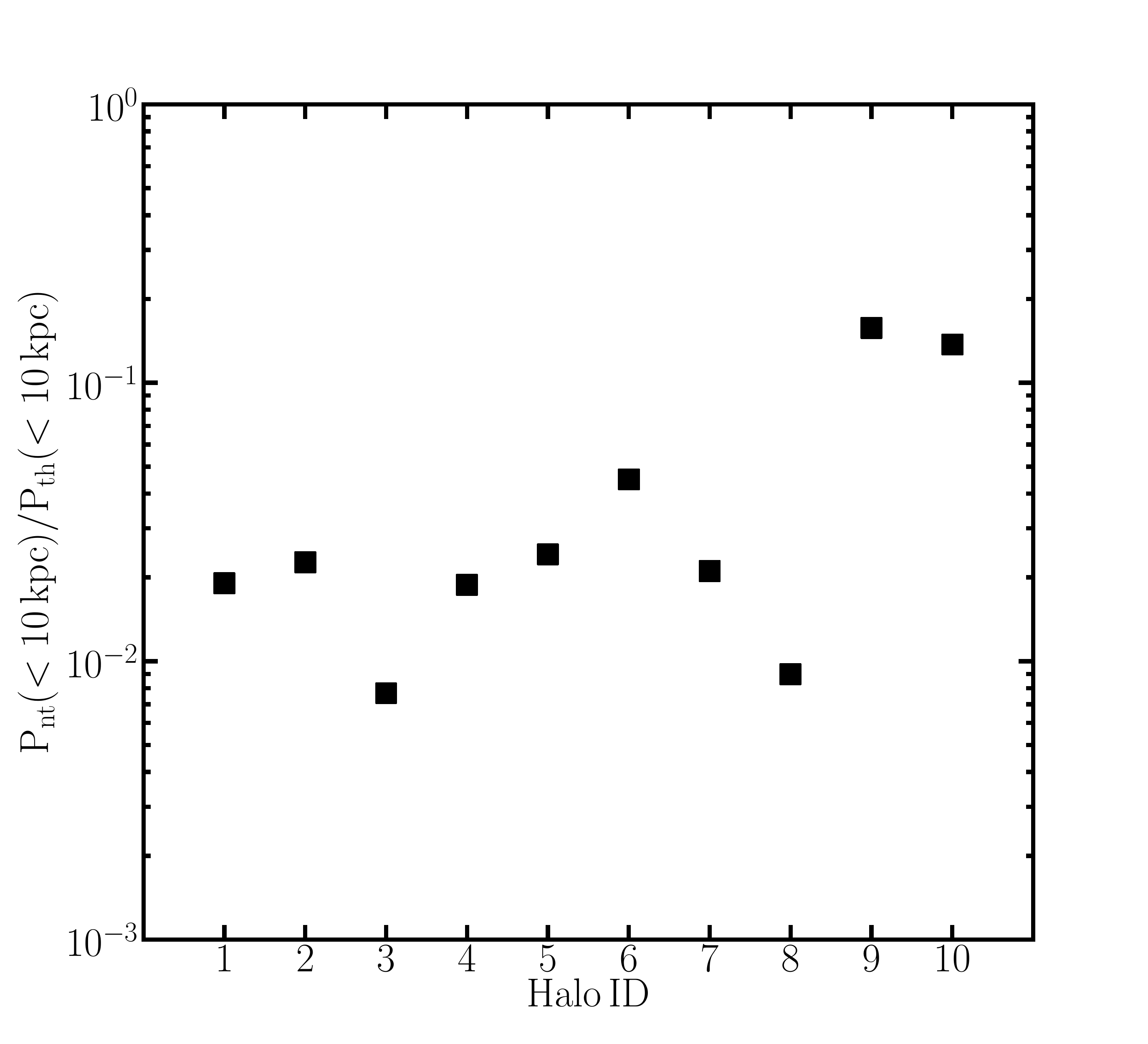}
\caption{\label{fig:central_bc} Ratio between central non-thermal pressure and central thermal pressure for the halos in our sample. Halos 1-5 are relaxed, halos 6-10 are not relaxed.} 
\end{figure}

\subsection{Non-thermal pressure from ICM velocity dispersion}\label{sec:ntp-vdisp}

In simulations, it is in principle possible to calculate an approximate estimate of the non-thermal pressure as a function of radius as
\begin{equation}\label{eq:pdisp}
 {\rm P_{nt}\approx P_{disp}= \rho\sigma_{ICM}^2,}
\end{equation}
where ${\rm \sigma_{ICM}}$ is the velocity dispersion in the gas motions. We used Equation~\ref{eq:pdisp} to estimate an approximate value for the non-thermal pressure in the region ${\rm r < 10}$ kpc, ${\rm P_{\rm nt}(r<10\hbox{ }kpc)}$, and compared its value to the thermal pressure within the same region ${\rm P_{\rm therm}(r<10\hbox{ }kpc)}$. 
The result of the comparison is shown in Figure~\ref{fig:central_bc}. For 8 of our halos the central non-thermal pressure is a few \% of the central thermal pressure, for the other two (unrelaxed) clusters the ration increases to $\sim$10 \%.

To assess the robustness of Equation~\ref{eq:pdisp} we also measure the full radial profile ${\rm P_{disp}(r)}$ and use it to perform yet another mass reconstruction:
\begin{equation}
 {\rm M_{disp}(<r)=-\frac{r^2}{G\rho(r)}\frac{d}{dr}[P_{therm} (r) + P_{disp}(r)]}
\end{equation}
All the mass estimators defined so far will be compared in the next Section.

\section{Results}\label{sec:results}

\begin{figure*}
\includegraphics[width=.99\textwidth]{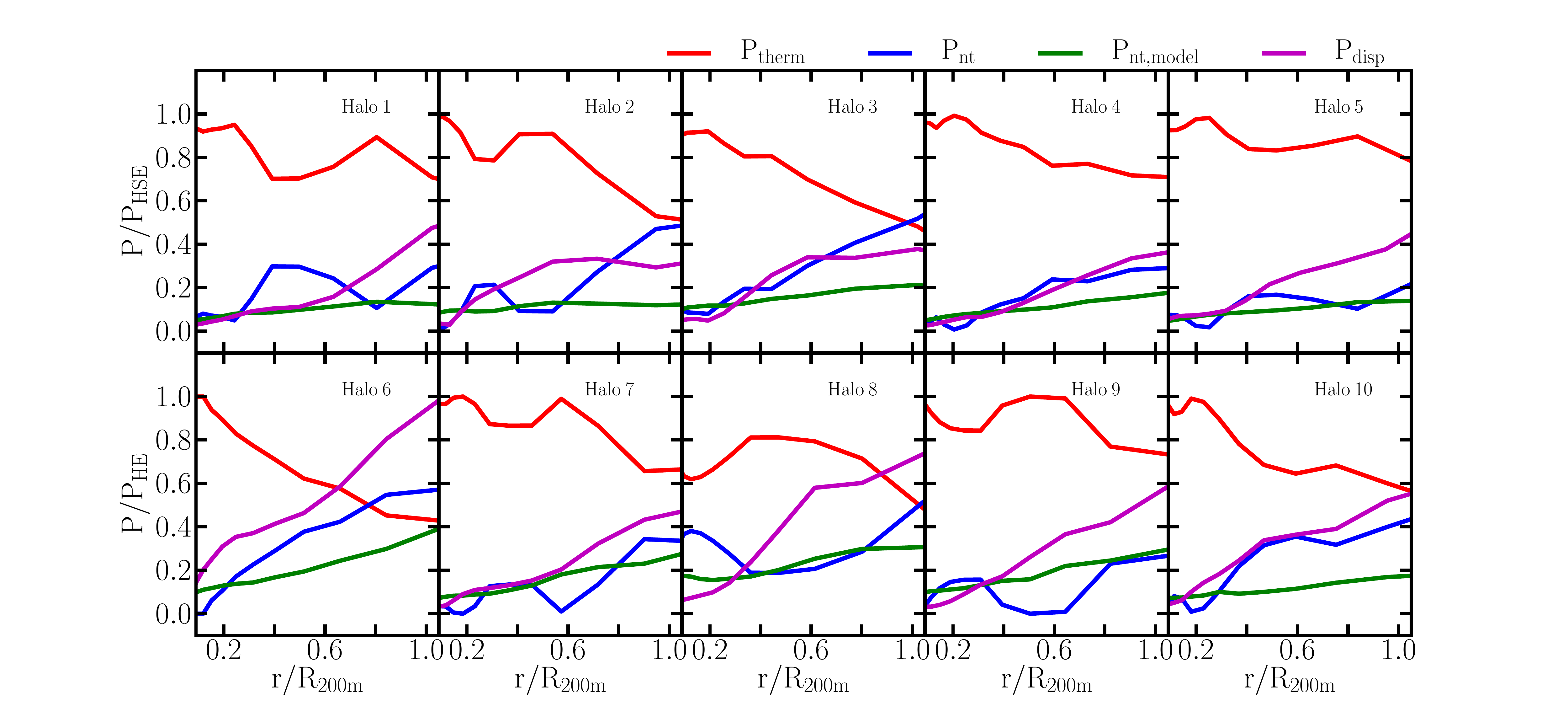}
\caption{\label{pressure_fractions} Fraction of pressure contributions with respect to the total pressure as a function of radius, under the assumption of spherical symmetry and hydrostatic equilibrium. Each panel represents one of the 10 halos analysed in this paper. Red solid lines represent the contribution from thermal pressure; blue solid lines represent the contribution from non-thermal pressure; green solid lines represent the empirical formula in Equation~\ref{eq:model_pnt}; magenta solid lines represent the approximate value for the non-thermal pressure ${\rm P_{disp}= \rho\sigma_{ICM}^2}$.}
\end{figure*}

We begin the presentation of our results by showing the ratio of different pressure contributions as a function of radius for our halos at redshift $z=0$ in  Figure~\ref{pressure_fractions}. The total pressure ${\rm P_{\rm tot}}$ used as reference in this figure is the one required to maintain hydrostatic equilibrium in each cluster (${\rm P_{HE}}$ in Equation~\ref{eq:ptot}). The thermal pressure profile is directly measured from the simulations. The non-thermal pressure is computed as ${\rm P_{nt}=1-P_{\rm therm}}$. Our results show significant halo-to-halo differences, but a general trend is evident in which thermal pressure dominates in the central regions of the clusters, with non-thermal pressure becoming increasingly relevant at large radii and constituting $\sim 20-40 \%$ of the total pressure. This result is in general agreement with the analysis performed by \cite{2015MNRAS.448.1020S} and \cite{2016arXiv160602293B}. In a few of the clusters there are peaks of ${\rm P_{nt}/P_{HE}}\sim 30-40 \%$ at a distance of a few hundred kpc from the center. Such objects typically have more disturbances in the density/velocity fields. 

The green line in Figure~\ref{pressure_fractions} represents the model for non-thermal pressure of Equation~\ref{eq:model_pnt} which aims at approximating the non-thermal pressure profile of the simulated clusters. The model reproduces the general trend of increasing ${\rm P_{nt}/P_{HE}}$ with cluster-centric radius, but is unable to reproduce the local fluctuations of non-thermal pressure seen in the actual profiles. Despite this issue, we stress that the goal of Equation~\ref{eq:model_pnt} is not to provide an exact measure of the non-thermal pressure at a given radius, but rather an empirical formula to improve the results of mass reconstructions of galaxy clusters. 

The magenta line in Figure~\ref{pressure_fractions} represents the ratio ${\rm P_{disp}/P_{HE}}$ as a function of radius, where ${\rm P_{disp}=\rho\sigma_{ICM}^2}$ dynamically probes non-thermal pressure. In most clusters, ${\rm P_{disp}}$ agrees with the other estimates of non-thermal pressure only in central regions and it is usually in excess at large radii. Better agreement between ${\rm P_{ nt}}$, ${\rm P_{nt,model}}$ and ${\rm P_{\rm disp}}$ can be appreciated in the relaxed clusters (IDs 1 to 5). 

\begin{figure*}
\includegraphics[width=.99\textwidth]{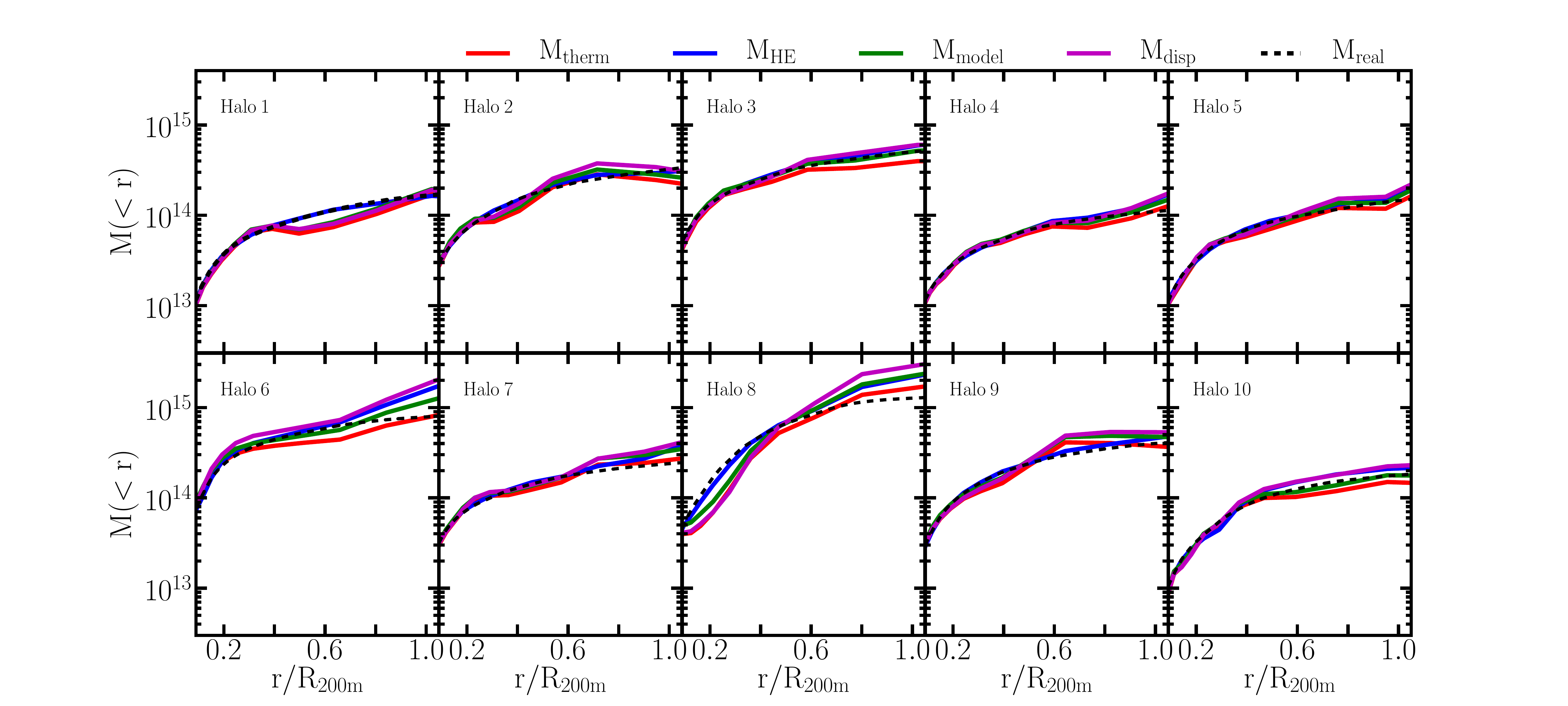}
\caption{\label{mass_profiles} Spherically averaged enclosed mass profiles as a function of radius. Each panel represents one of the 10 halos analysed in this paper. Black dashed lines represent the real mass profile directly measured from the simulations; red solid lines represents the mass reconstruction achieved by assuming spherical symmetry, hydrostatic equilibrium and by considering only the thermal pressure; blue solid lines are the mass profiles obtained by assuming spherical symmetry, hydrostatic equilibrium and the contribution from thermal and non-thermal pressure (this profile can be only measured explicitly in simulations); green solid lines are the theoretical mass reconstructions assuming spherical symmetry, hydrostatic equilibrium and the contribution of thermal and non-thermal pressure following the empirical model in Equation~\ref{eq:model_pnt}; magenta solid lines represent mass reconstructions assuming spherical symmetry, hydrostatic equilibrium and the contribution of thermal and non-thermal pressure, with the latter approximated by ${\rm P_{\rm nt}\approx P_{disp}= \rho\sigma_{ICM}^2}$.}
\end{figure*}

The quality of the different approaches to mass reconstruction discussed in Section~\ref{sec:formalism} can be assessed by considering Figure~\ref{mass_profiles} which shows the spherically averaged mass profile ${\rm M_{real}}$ directly measured from each simulated cluster compared to the result of several mass reconstructions techniques. If one assumes spherical symmetry and hydrostatic equilibrium, the `true'  hydrostatic mass ${\rm M_{HE}}$ can be estimated by measuring the radial component of the gravity field from the simulations and then using Equation~\ref{eq:mass_hse}; it is important to stress that ${\rm M_{HE}}$ can only be measured from simulations that provide the value of the gravity field. Differences between ${\rm M_{real}}$ and ${\rm M_{HE}}$ represent the combined effect of hydrostatic bias and deviations from spherical symmetry. These differences are typically noticeable in this plot at large radii ${\rm r/R_{200m}>0.6-0.7}$, but also at smaller radii in some of the more dynamically disturbed systems (e.g. halos 8 and 10). 
It appears that the assumption of hydrostatic equilibrium yields accurate mass reconstructions for 7 out of 10 clusters in the range ${\rm 0.1<r/R_{200m}<0.7}$. 
However, for relaxed clusters, the agreement between ${\rm M_{\rm HE}}$ and ${\rm M_{real}}$ is excellent in the range ${\rm 0.1<r/R_{200m}<1.1}$ (see quantitative comparison in Subsection~\ref{sec:quant}). 

The most important comparison in Figure~\ref{mass_profiles} is the one between ${\rm M_{real}}$ and ${\rm M_{therm}}$. The latter is the most basic mass reconstruction that an observer with access to density and thermal pressure measurements can perform, but it neglects the contribution from non-thermal pressure  (Equation~\ref{eq:mtherm}). Figure~\ref{mass_profiles} confirms that ${\rm M_{therm}}$ is a biased estimator both with respect to the real mass ${\rm M_{real}}$ and to the total hydrostatic mass ${\rm M_{HE}}$. As expected, ${\rm M_{therm}}$ typically underestimates the mass of the cluster. 

Figure~\ref{mass_profiles} shows that the theoretical model of Equation~\ref{eq:mass_theory}, ${\rm M_{model}}$. In this case, the contribution of non-thermal pressure to the equation of hydrostatic equilibrium is taken into account following Equation~\ref{eq:model_pnt}. ${\rm M_{model}}$ provides a more accurate hydrostatic mass reconstruction than ${\rm M_{therm}}$, i.e. ${\rm M_{model}}$ is closer to the `true' hydrostatic mass, ${\rm M_{HE}}$. 

\subsection{Quantitative assessment of mass reconstruction biases}\label{sec:quant}

\begin{figure}
\includegraphics[width=.49\textwidth]{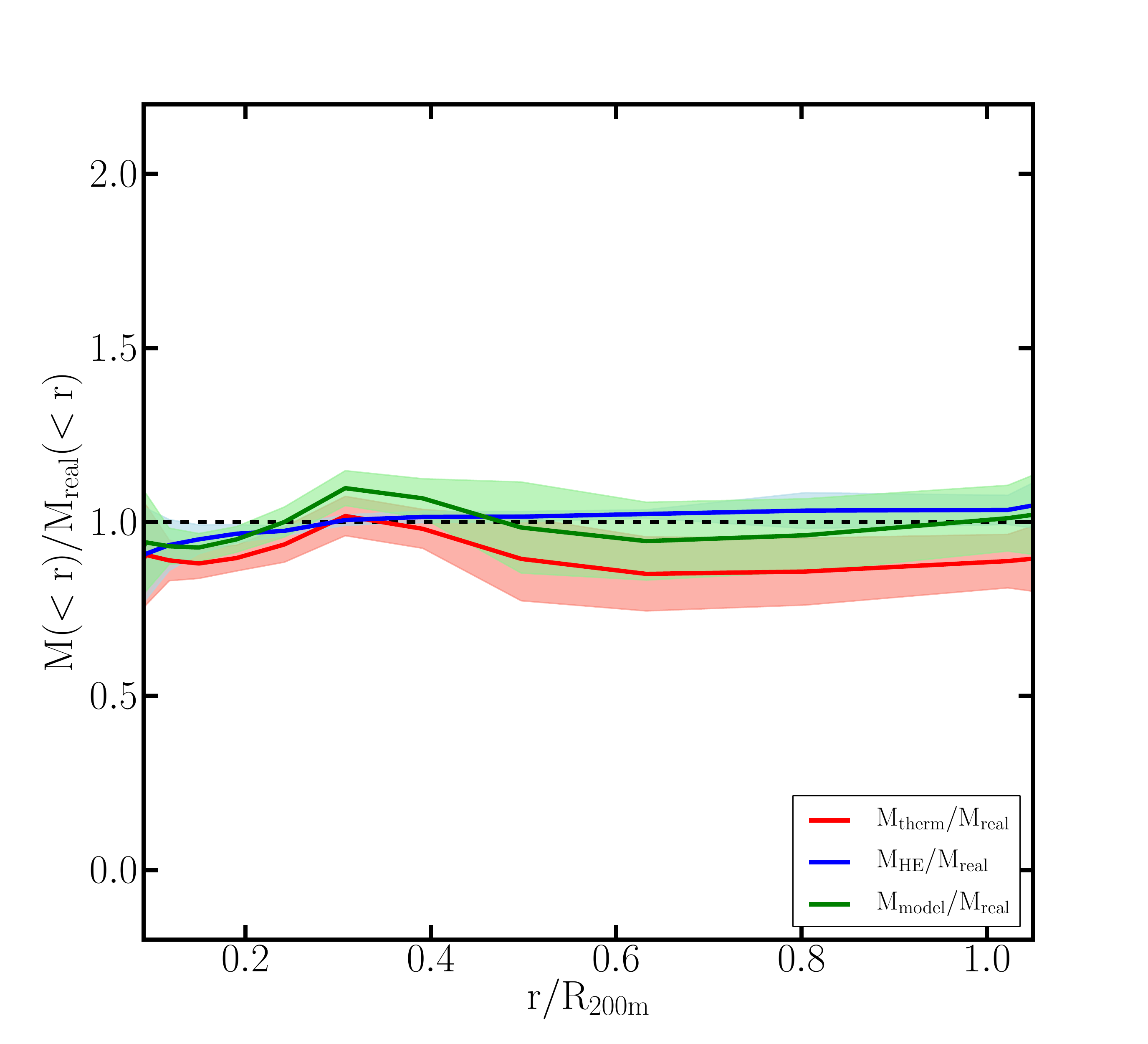}
\caption{\label{mass_bias} Average mass reconstruction bias ${\rm M(<r)/M_{real}(<r)}$ for relaxed clusters as a function of radius for different mass estimators. The solid lines represent the average among the 10 simulated halos and the shaded areas represent the halo-to-halo 1-$\sigma$ scatter.} 
\end{figure}

A more quantitative view of the situation is provided by Figure~\ref{mass_bias} that shows the average mass reconstruction bias ${\rm M(<r)/M_{real}(<r)}$ as a function of radius {\itshape for the relaxed clusters}. This plot shows that all the mass estimators we considered are $\sim10\%$ biased at radii ${\rm r<0.2R_{500}}$; within this region all clusters are relatively unrelaxed (see Figure 1), with the implication that hydrostatic equilibrium is a bad approximation. At larger radii, which are more relevant for cosmological applications, ${\rm M_{HE}}$ is the least biased reconstruction with accuracy $\sim 5\%$. This estimator would be useful only if both thermal and non-thermal pressure could be explicitly measured. Unfortunately, this is only feasible in clusters for which high quality data is available \citep[e.g.][]{2013MNRAS.428.2241S, 2015ApJ...806..207U}, but not in general. 

Figure~\ref{mass_bias} shows that ${\rm M_{therm}}$ is always underestimating the real mass by $10-20 \%$ at ${\rm r>0.5R_{500}}$. However, ${\rm M_{model}}$, the theoretical model we propose in this paper (Equation~\ref{eq:mass_theory}) is very successful at correcting for the bias in ${\rm M_{therm}}$. On average, ${\rm M_{model}}$ provides a mass reconstruction that is always very close to the `true' hydrostatic mass ${\rm M_{HE}}$. This result is potentially very useful, because it shows that it is possible to weaken at least one source of bias in cluster mass reconstructions: the lack of knowledge on non-thermal pressure from turbulent and bulk motions in the ICM.

\section{Summary and Discussion}\label{sec:conclusion}

We reviewed the basics of cluster mass reconstruction under the assumption of hydrostatic equilibrium. If deviations from spherical symmetry and contributions from non-thermal pressure are neglected, it is possible to use the equation of hydrostatic equilibrium to yield a reconstruction of a galaxy cluster mass profile. This approach has been widely used in the literature by combining constraints on the ICM thermal pressure from detection of the thermal SZ effect in galaxy clusters \citep{2010A&A...517A..92A, 2013ApJ...768..177S, 2013MNRAS.430.1344O, 2015A&A...583A.111R} and constraints from X-ray observations of galaxy clusters \citep{2010MNRAS.402...65S, 2010PASJ...62..371H, 2013SSRv..177..119E}. 

The question of whether additional contribution to pressure support in galaxy clusters has been addressed by many authors \citep{2007MNRAS.378..385P, 2009ApJ...705.1129L, 2010ApJ...711.1033Z, 2012ApJ...758...74B, 2013ApJ...771..102F, 2013MNRAS.432..404M,  2015ApJ...808..176A}. In this paper, we extended this line of research by characterizing the contribution from non-thermal pressure support in a sample of 10 galaxy clusters simulated with  the cosmological AMR code {\sc ramses} \citep{2002A&A...385..337T}. With our analysis we were also able to quantify the combined bias of the assumption of hydrostatic equilibrium and spherical symmetry. 

We find that the traditional hydrostatic mass reconstruction that only considers of thermal pressure in the ICM underestimated the cluster mass by $10-20 \%$ at radii ${\rm 0.1R_{500}<r<2R_{500}}$, in agreement with recent results in the literature \citep{2016MNRAS.455.2936S, 2016arXiv160602293B}.

The contribution from non-thermal pressure to the support of the ICM against gravity is significant and typically increases with radius, with a  maximum contribution of $20-40 \%$ of the total pressure at radii ${\rm R_{500}<r<2R_{500}}$. We showed that adding this contribution is important to remove the bias present in the traditional hydrostatic mass reconstruction method. 

One of the outstanding issues for the determination of the contribution from non-thermal pressure in clusters is that it is hard to constrain from observations \citep[however, see][]{2010ApJ...711.1033Z}. In this paper, we use our simulations to  calibrate a formula for the non-thermal pressure as a function of ICM density which is used to remove the mass reconstruction bias. The use of such a model for the non-thermal pressure  provides an improved hydrostatic cluster mass estimator. 

The most important caveat about the method we propose is that even if non-thermal pressure is accounted for, the assumptions of hydrostatic equilibrium and spherical symmetry implicitly carry with them a 5-10 \% mass bias. These effects are more difficult to study and to remove \citep[e.g.][]{2015MNRAS.448.1644S, 2016MNRAS.460..844M} and we defer the analysis of these issues to future work. 

An additional caveat is that there are additional sources of pressure support that are not included in our simulations (e.g. magnetic fields and cosmic ray pressure) whose relevance for cluster mass modeling needs to be assessed. 

Finally, the spatial resolution $\sim 1$ kpc/h of our simulations does not allow us to explicitly test whether a full the turbulent cascade is achieved. In this case turbulence is expected to dissipate faster than in nature. \cite{2012A&A...544A.103V} performed a detailed analysis of turbulence in simulated clusters and identified resolution effects on scales $\sim 10-20$ kpc. Therefore, we conclude that our results only provide a {\it lower limit} to the contribution of turbulence to pressure balance in clusters and thus a lower limit to the mass bias expected by neglecting this source of pressure support. This issue will be much better addressed by the next generation of cosmological zoom-in simulations which will have $\sim 10$ times better resolution and will include more physical processes. 
Fortunately, our simulations have enough resolution to capture the effect of the other large contribution to non-thermal pressure, bulk motions, which typically play an important role at large cluster-centric distances ($r>500$ kpc, \cite{2015MNRAS.448.1020S}). 

In conclusion, the approach we propose for mass reconstruction constitutes a significant step forward with respect to traditional methods based on hydrostatic equilibrium. This improvement will be relevant for better mass  calibration of cluster scaling relations in the era of large surveys for precision cosmology (e.g. Euclid, LSST, DESI, eROSITA). Re-calibration of cluster scaling relations with the proposed method using archival data (e.g. from HST, ACT, SPT, Planck) and data from future X-ray surveys (eROSITA, ATHENA) will significantly improve the precision of cosmological constraints coming from cluster studies (e.g. cluster mass function). 

\section*{Acknowledgments}
D.M. acknowledges support from the Swiss National Science Foundation (SNSF) through the SNSF Early.Postdoc and Advanced.Postdoc Mobility Fellowships. This work was supported by a grant from the Swiss National Supercomputing Centre (CSCS).

\bibliography{paper} 

\label{lastpage}

\end{document}